\documentclass[conference]{IEEEtran}
\IEEEoverridecommandlockouts

\usepackage{cite}
\usepackage{amsmath,amssymb,amsfonts}
\usepackage{algorithmic}
\usepackage{textcomp}

\usepackage{xcolor}
\ifCLASSINFOpdf
  \usepackage[pdftex]{graphicx}
  \graphicspath{{./image/}}
  \DeclareGraphicsExtensions{.pdf,.png,.jpeg,.jpg}
\else
  \usepackage[dvips]{graphicx}
  \graphicspath{{./image/}}
  \DeclareGraphicsExtensions{.eps}
\fi
\usepackage{booktabs}
\usepackage{array}
\usepackage{enumitem}
\setlist[description]{style=nextline}
\newcommand{\etal}{\textit{et~al.}}

\def\BibTeX{{\rm B\kern-.05em{\sc i\kern-.025em b}\kern-.08em
    T\kern-.1667em\lower.7ex\hbox{E}\kern-.125emX}}
\begin{document}

\title{Indoor Space Authentication by ISS-based Keypoint Extraction from 3D Point Clouds}

\author{\IEEEauthorblockN{Yuki Yamada}
\IEEEauthorblockA{\textit{Kyoto University} \\
yamada@inet.media.kyoto-u.ac.jp}
\and
\IEEEauthorblockN{Daisuke Kotani}
\IEEEauthorblockA{\textit{Kyoto University} \\
kotani@media.kyoto-u.ac.jp}
\and
\IEEEauthorblockN{Kota Tsubouchi}
\IEEEauthorblockA{\textit{LY Corporation} \\
ktsubouc@lycorp.co.jp}
\and
\IEEEauthorblockN{Hidehito Gomi}
\IEEEauthorblockA{\textit{LY Corporation} \\
hgomi@lycorp.co.jp}
\and
\IEEEauthorblockN{Yasuo Okabe}
\IEEEauthorblockA{\textit{Kyoto University} \\
okabe@media.kyoto-u.ac.jp}
}

\maketitle

\begin{abstract}
We propose ISS-RegAuth, a lightweight indoor space authentication framework that authenticates a user by comparing LiDAR captures of personal rooms.
Prior work processes every point in the cloud, where planar surfaces such as walls and floors dominate similarity calculations, causing latency and potential privacy exposure.
In contrast, ISS-RegAuth retains only 1–2\% of Intrinsic Shape Signatures (ISS) keypoints, computes their Fast Point Feature Histograms, and performs RANSAC and ICP on this sparse set.
On 100 ARKitScenes pairs, this approach reduces the equal-error rate from 0.02 to 0.00, cuts processing time by 20\%, and lowers transmitted data to 2.2\% of the original.
These results show that keypoint-based sparse representation can make privacy-preserving, edge-deployable indoor space authentication practical.
As an early step, this work opens a path toward device-independent authentication and account-recovery mechanisms that rely on users’ physical environments.
\end{abstract}

\begin{IEEEkeywords}
Indoor Space Authentication, 3D Point Cloud, ISS Algorithm, Privacy, Account Recovery
\end{IEEEkeywords}

\section{Introduction}
\begin{figure}[t]
    \centering
    \includegraphics[width=0.9\linewidth]{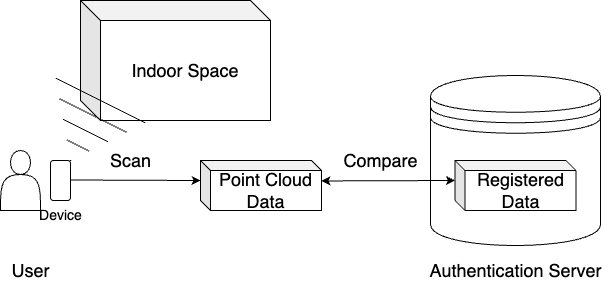}
    \caption{Indoor space authentication concept. Users enroll by scanning a private room and later re-scan it for verification.}
    \label{fig:concept}
\end{figure}

Indoor space authentication offers a novel approach to identity verification by leveraging the unique 3D geometry of a user's physical environment (Fig.~\ref{fig:concept}).
This approach anchors a user's identity to their physical environment rather than device ownership, introducing a new authentication factor based on ``what space the user possesses.''

Suzuki \etal \cite{suzuki2023} pioneered LiDAR-based indoor space authentication, demonstrating that 3D spatial geometry captured by LiDAR can serve as a reliable authentication factor.
However, LiDAR-based approaches face unique challenges: users must rescan their environment with each authentication attempt, leading to variations in capture range and sensor alignment.
This requires robust matching algorithms that can handle such variations to correctly identify the same room.
Moreover, transmitting 3D structure may expose privacy-sensitive objects and layout information about the user's living space, yet existing work has not adequately addressed these concerns.

To address these limitations, we propose ISS-RegAuth, a framework that shifts from dense matching to sparse representation.
Instead of transmitting raw scans, our method extracts only 1–2\% of points as Intrinsic Shape Signature (ISS) keypoints and performs authentication based solely on these sparse features.
On 100 ARKitScenes pairs, ISS-RegAuth achieves:
\begin{description}
    \item[(i)] 98\% reduction in transmitted and stored data,
    \item[(ii)] 20–30\% faster verification on a MacBook Air (M3),
    \item[(iii)] $\mathrm{EER}=0.00$ by using 2\% keypoints (on this dataset)
\end{description}
Furthermore, our experiments demonstrate robustness to significant variations in capture range, correctly matching rooms even when scanning coverage differs substantially.
These results provide a key step toward making indoor space authentication practical for real-world applications, particularly for account recovery scenarios where users can regain access with a newly purchased device, for instance.

\section{Related Work}

Authentication systems that use indoor location have been explored through other sensing modalities.
Portnoi and Shen \cite{locauth} proposed location-enabled authentication using Wi-Fi signal characteristics, while Li \etal \cite{sencs} verified user presence in a room by observing human actions.
While various types of indoor spatial information have been explored for authentication, this paper focuses on LiDAR-based approaches that utilize 3D geometric structure.

LiDAR sensors emit laser pulses, measure their return time, and convert the distances into dense 3D point clouds that can be stored in standard formats (e.g., .ply or .pcd).
Modern iPhones and iPad Pro devices include short-range LiDAR units and inertial sensors, allowing everyday users to capture indoor scenes within approximately five meters.

Most 3D registration pipelines follow a common recipe.
First, local geometric descriptors such as Fast Point Feature Histograms (FPFH) \cite{fpfh} are computed for each point.
Second, correspondences between similar descriptors are matched and a rigid transform is estimated using robust methods like RANSAC \cite{ransac}.
Third, the alignment is refined through Iterative Closest Point (ICP) \cite{icp}.

Processing every point in a dense cloud is computationally expensive and can expose unnecessary geometric details.
To mitigate this, keypoint detectors such as Intrinsic Shape Signatures (ISS) \cite{iss} have been introduced to identify salient points that best represent structural features like corners and edges.

Suzuki \etal \cite{suzuki2023} applied the FPFH-RANSAC-ICP registration pipeline to indoor space authentication using every down-sampled point.
Although effective, their approach required transmitting thousands of points per scan, causing planar regions such as walls and floors to dominate similarity scores and resulting in significant privacy exposure.

Public datasets including ScanNet++ \cite{scannetpp} and ARKitScenes \cite{arkitscenes} have made quantitative evaluation feasible.
Among these, ARKitScenes is particularly suitable for this study because it provides over 1,600 high-quality captures from iPad Pro devices, including multiple scans per room recorded under realistic conditions.

\section{ISS-RegAuth; Indoor Space Authentication by ISS-based Keypoint Extraction}
\subsection{Authentication Procedure}
Indoor space authentication operates in two phases:

\textbf{Enrollment:}
The user scans a personal indoor space (e.g., their own room) using a LiDAR-capable device.
The captured 3D point cloud is processed and registered on the server as a template associated with the user's account.

\textbf{Verification:}
When authentication is required, the user rescans the same space with a LiDAR-capable device.
The probe scan is processed and transmitted to the server, which computes the geometric similarity between the template and probe.
The user is authenticated if the similarity exceeds a predefined threshold.

\subsection{ISS-RegAuth Processing Pipeline}

\textbf{Enrollment Phase:}
\begin{enumerate}
    \item \textbf{Pre-processing:} Raw point clouds are voxel down-sampled and smoothed via PCA-based normal estimation. Radius outlier removal eliminates isolated points caused by sensor noise.
    \item \textbf{ISS keypoint extraction:} The ISS algorithm identifies salient points representing structural features such as corners and edges, based on eigenvalue ratios of the local covariance matrix.
    \item \textbf{Descriptor computation:} FPFH descriptors are calculated only for the extracted keypoints, yielding rotation-invariant geometric descriptors.
    \item These sparse keypoints and descriptors are transmitted to the server and stored as a template.
\end{enumerate}

\textbf{Verification Phase:}
\begin{enumerate}
    \item \textbf{Device-side processing:} The user rescans the space, and the sparse data processed identically is transmitted.
    \item \textbf{Registration (server-side):} RANSAC estimates a coarse rigid transformation between the stored template and the newly captured scan, followed by ICP refinement for precise alignment.
    \item \textbf{Similarity computation (server-side):} The ratio between matched template keypoints and the total number of template keypoints is computed. The user is authenticated if similarity exceeds a predefined threshold.
\end{enumerate}

Prior work by Suzuki \etal~\cite{suzuki2023} computes descriptors for all down-sampled points and performs registration on the entire point cloud.
This causes large planar surfaces such as walls and floors to dominate similarity calculations.
In contrast, ISS-RegAuth extracts only salient keypoints that capture structural features, emphasizing furniture-scale geometry that better distinguishes individual rooms while reducing both computation and data exposure.

\begin{figure}[t]
    \centering
    \begin{minipage}[t]{0.48\linewidth}
        \centering
        \includegraphics[width=\linewidth]{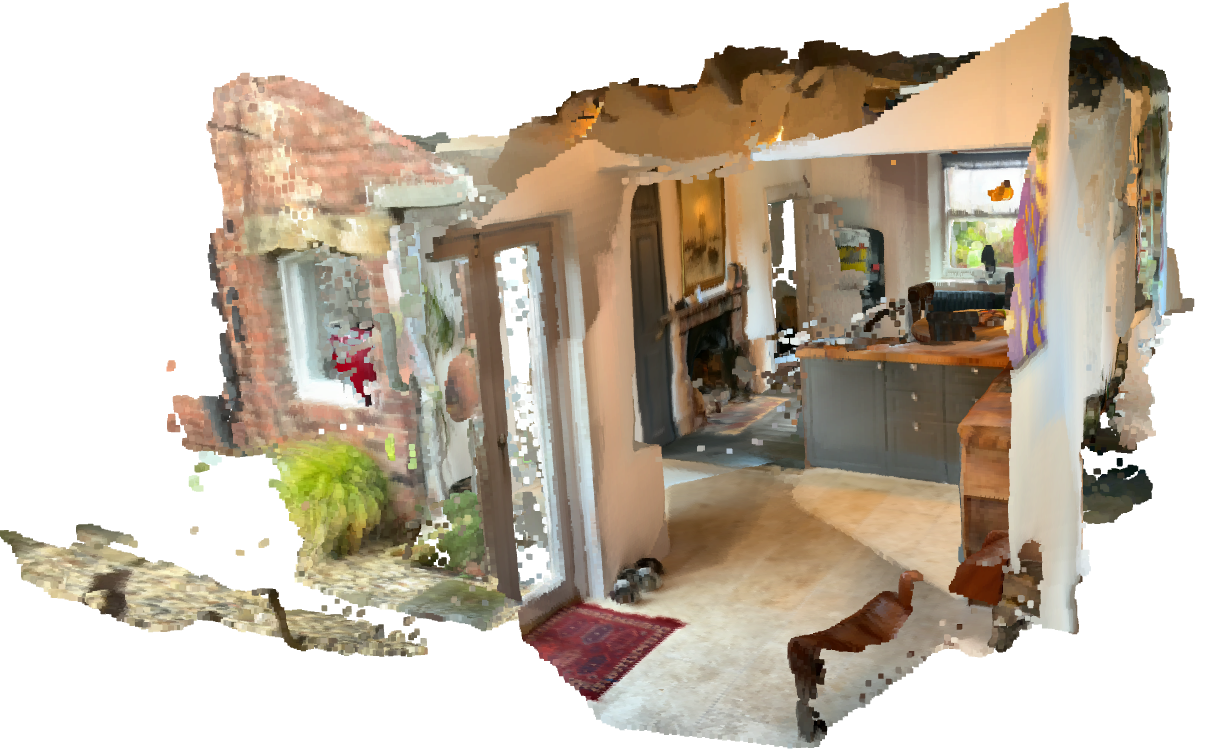}
    \end{minipage}
    \hfill
    \begin{minipage}[t]{0.48\linewidth}
        \centering
        \includegraphics[width=\linewidth]{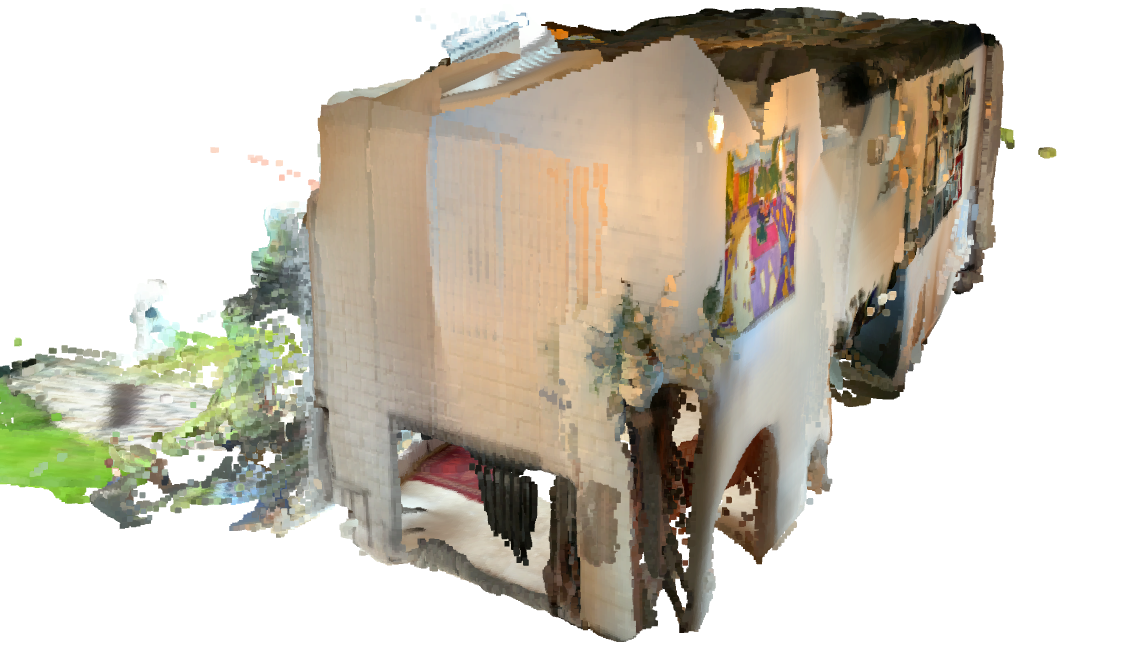}
    \end{minipage}
    \caption{Example pair of ARKitScenes captures from the same room used for enrollment (left) and probing (right).}
    \label{fig:same-space-pair}
\end{figure}

\begin{figure}[t]
    \centering
    \includegraphics[width=0.9\linewidth]{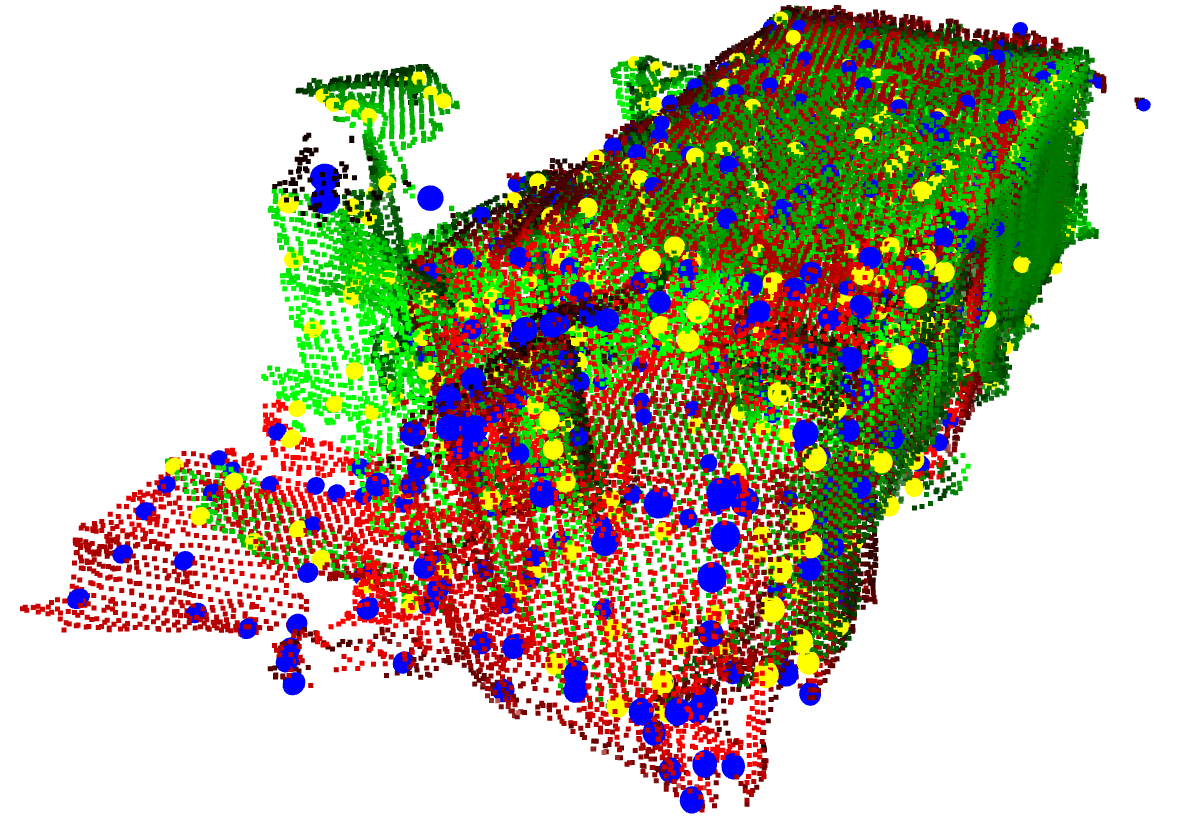}
    \caption{ISS-RegAuth aligns only the salient keypoints (yellow/blue spheres) on top of the down-sampled clouds (green/red).}
    \label{fig:registration-example}
\end{figure}

Figures~\ref{fig:same-space-pair} and~\ref{fig:registration-example} illustrate how the sparse keypoints capture furniture-scale geometry while ignoring large planar regions such as floors and walls.

\section{Evaluation}
\subsection{Experimental Setup}
We evaluate ISS-RegAuth using the ARKitScenes dataset \cite{arkitscenes}, which includes over 1,600 real-world indoor captures from iPad Pro devices.
To emulate realistic verification scenarios, we randomly select 50 pairs of scans from the same space and 50 pairs from different spaces, allowing for natural variations in camera trajectories and partial overlaps.
All experiments are conducted in Python 3.12.10 using Open3D 0.19.0 on a MacBook Air (M3, 16 GB RAM), representing a lightweight edge-verification environment.

We compare three configurations: the baseline method \cite{suzuki2023}, ISS-RegAuth with approximately 2\% keypoints, and ISS-RegAuth with approximately 1\% keypoints.
The following parameters are used across all experiments:

\textbf{Pre-processing:} Voxel down-sampling at 0.1~m resolution, PCA-based normal estimation, and radius outlier removal.

\textbf{ISS keypoint extraction:} Salient radius = 0.2~m, non-maximum radius = 0.2~m (for 2\% keypoints) or 0.3~m (for 1\% keypoints), eigenvalue ratio thresholds $\gamma_{21}=\gamma_{32}=0.95$, minimum neighbors = 5.

\textbf{Registration:} RANSAC with correspondence distance = 0.2~m, maximum iterations = 1,000,000, confidence = 99.9\%. ICP with maximum iterations = 50 and distance threshold = 0.2~m.

Pre-processing and registration parameters are kept consistent across all methods to isolate the impact of keypoint selection.

\subsection{Evaluation Metrics}
We evaluate three aspects of system performance.
\textbf{Accuracy} is measured as $1-\mathrm{EER}$ when the false rejection and false acceptance rates are equal, representing authentication reliability.
\textbf{Processing time} is the average end-to-end time per pair on the server side, from point-cloud pre-processing through similarity computation, indicating computational practicality.
\textbf{Data reduction rate} is the ratio of points not transmitted to the server relative to using the full point cloud, representing how much spatial detail remains on the device.

\subsection{Results}
\begin{figure}[t]
    \centering
    \includegraphics[width=0.9\linewidth]{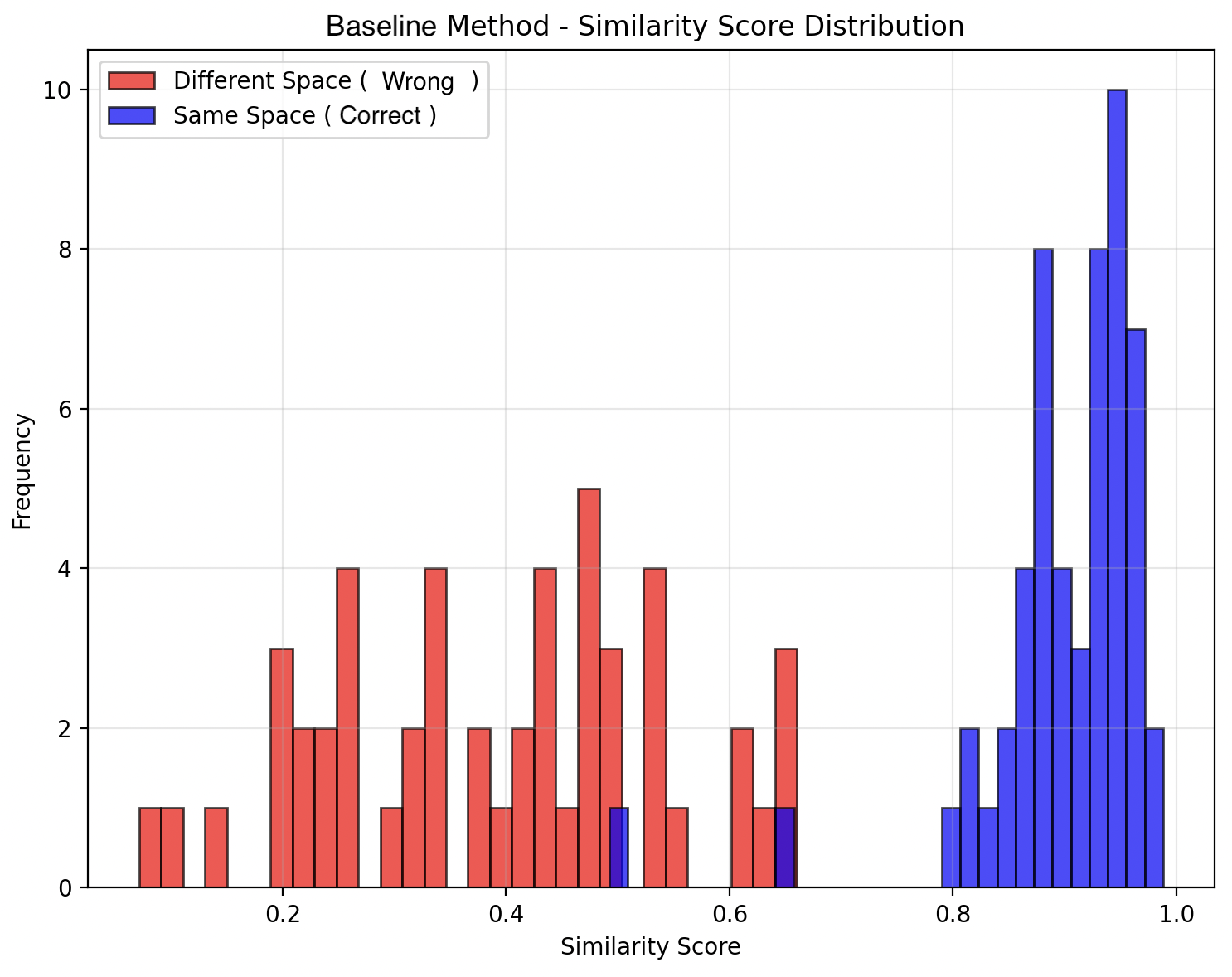}
    \vspace{2mm}
    \includegraphics[width=0.9\linewidth]{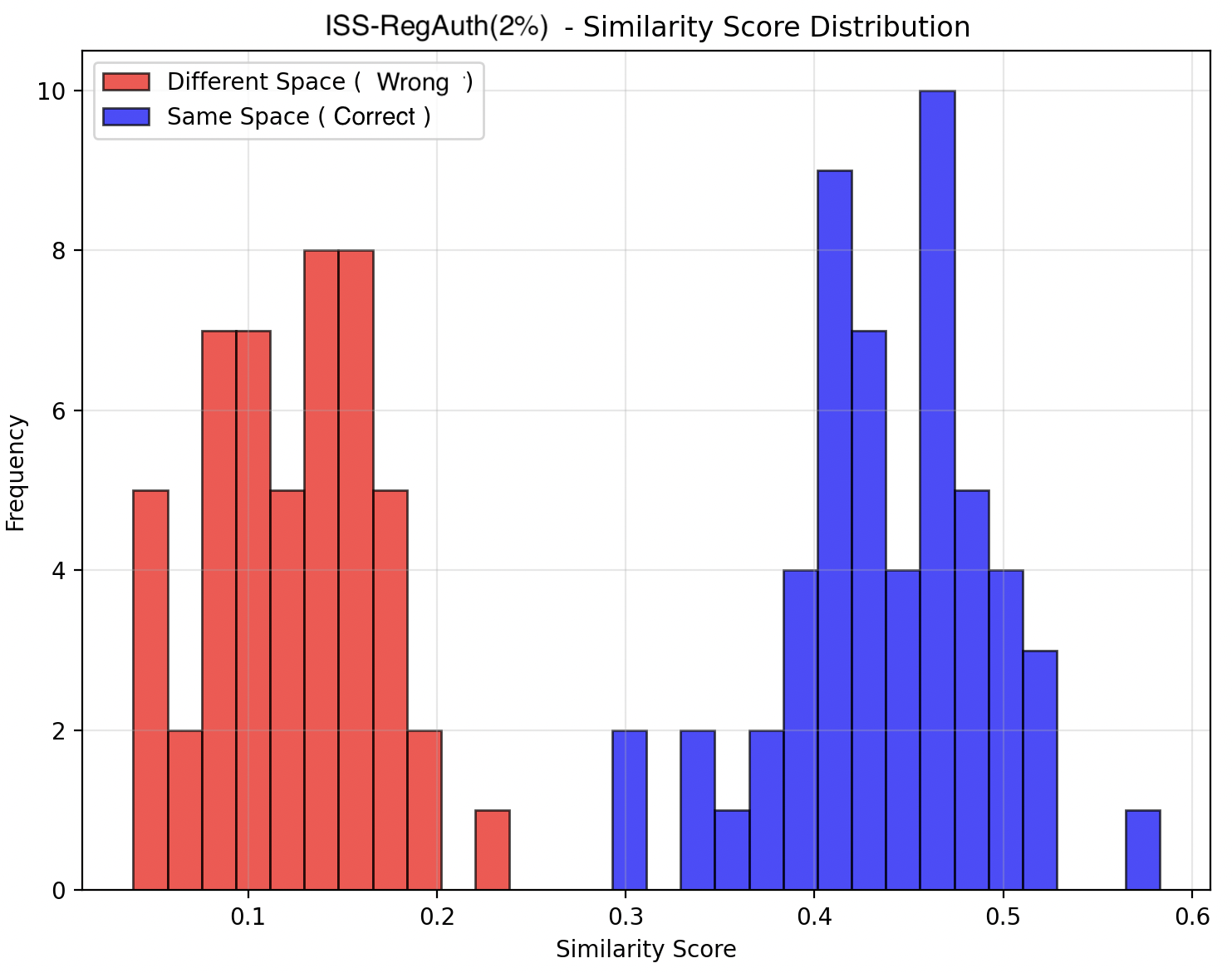}
    \caption{Similarity-score distributions for (top) the baseline~\cite{suzuki2023} and (bottom) ISS-RegAuth (2\%).}
    \label{fig:score-distribution}
\end{figure}

\textbf{Quantitative Analysis:}
Table~\ref{tab:performance} summarizes the performance comparison across three configurations.
Figure~\ref{fig:score-distribution} shows that ISS-RegAuth tightens the separation between correct and wrong pairs by suppressing high scores for geometrically similar yet distinct rooms, addressing the baseline's tendency to over-match based on planar surfaces.
The 2\% keypoint configuration achieves the best balance across all metrics, while the 1\% configuration trades accuracy for further reductions in computation and data transmission.

\begin{table}[t]
    \caption{Performance comparison on 100 ARKitScenes pairs.}
    \label{tab:performance}
    \centering
    \begin{tabular}{lccc}
        \toprule
        Method & Accuracy & Time (s) & Data reduction rate \\
        \midrule
        Suzuki \etal~\cite{suzuki2023} & 98\% & 3.58 & 0\% \\
        ISS-RegAuth (2\%) & 100\% & 2.70 & 97.8\% \\
        ISS-RegAuth (1\%) & 93\% & 2.45 & 98.9\% \\
        \bottomrule
    \end{tabular}
\end{table}

\textbf{Qualitative Analysis:}
Two representative cases highlight the robustness and privacy benefits of ISS-RegAuth.

\begin{figure}[t]
    \centering
    \includegraphics[width=0.9\linewidth]{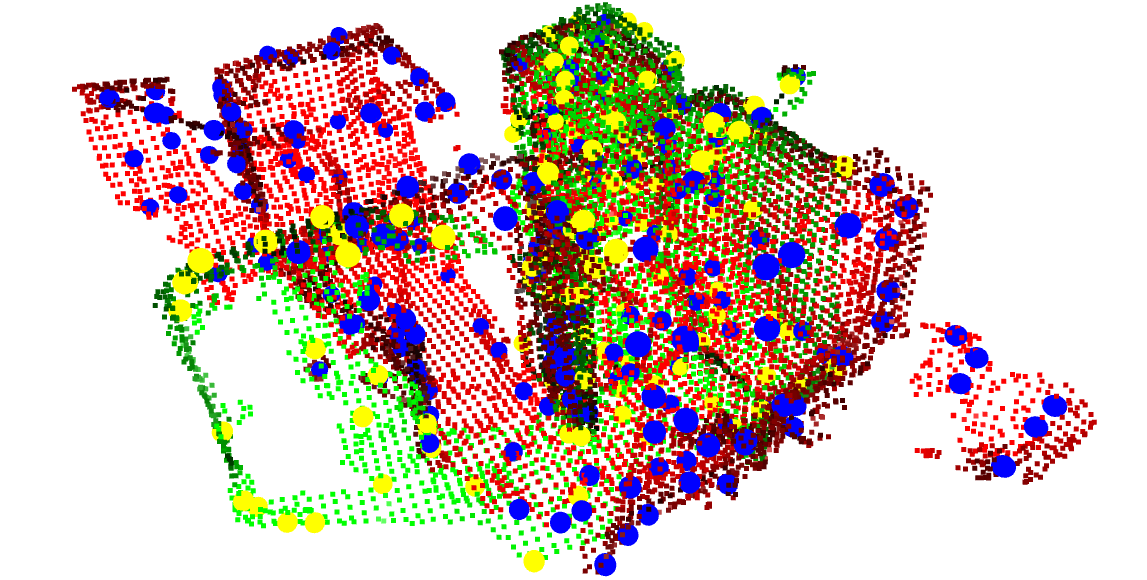}
    \caption{Example of enrollment and probe scans capturing partially different regions of the same room. ISS-RegAuth maintains consistent furniture-level correspondences despite limited overlap.}
    \label{fig:same-rooms}
\end{figure}

(1) Fig.~\ref{fig:same-rooms} shows a case where enrollment and probe scans capture partially different regions of the same room.
The baseline often fails in such scenarios due to limited wall overlap, whereas ISS-RegAuth successfully identifies consistent furniture-level correspondences,
yielding high similarity scores that correctly authenticate the user.

\begin{figure}[t]
    \centering
    \begin{minipage}[t]{0.48\linewidth}
        \centering
        \includegraphics[width=\linewidth]{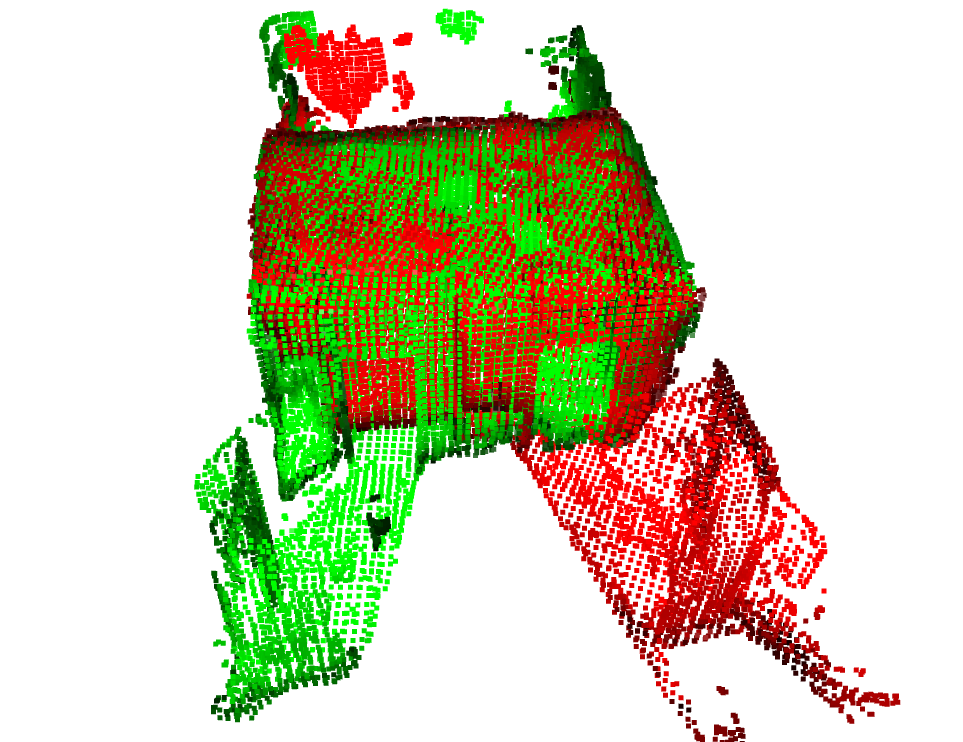}
    \end{minipage}
    \hfill
    \begin{minipage}[t]{0.48\linewidth}
        \centering
        \includegraphics[width=\linewidth]{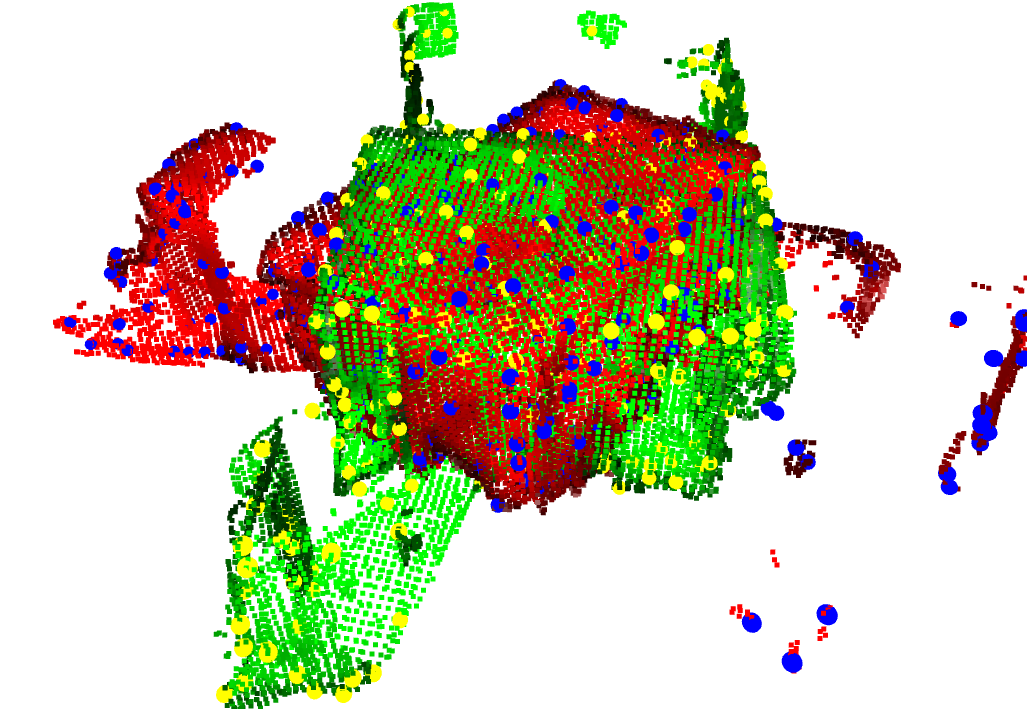}
    \end{minipage}
    \caption{Different rooms with deceptively similar rectangular layouts. The baseline achieves a high similarity (0.66) because walls align, whereas ISS-RegAuth rejects the pair (0.12) due to inconsistent keypoint correspondences.}
    \label{fig:similar-rooms}
\end{figure}

(2) For distinct rooms with similar rectangular layouts, the baseline occasionally produces false positives near 0.65 because planar walls align geometrically.
In contrast, ISS-RegAuth suppresses such cases (= 0.12 similarity) since sparse keypoints on movable objects rarely match (Fig. \ref{fig:similar-rooms}).

Overall, these results demonstrate that by de-emphasizing large planar structures and focusing on object-level geometry, ISS-RegAuth achieves practical indoor space authentication with improved discriminability and enhanced privacy.

\section{Discussion}
\subsection{Threat Model and Privacy}
We consider an adversary who gains access to stored templates on the server, either through a data breach or insider attack.
From these templates, the adversary may attempt to infer sensitive information about the user's living space, such as furniture arrangement, room layout, and functional areas (e.g., bedroom, bathroom), thereby violating the user's privacy.

\begin{figure}[t]
    \centering
    \begin{minipage}[t]{0.48\linewidth}
        \centering
        \includegraphics[width=\linewidth]{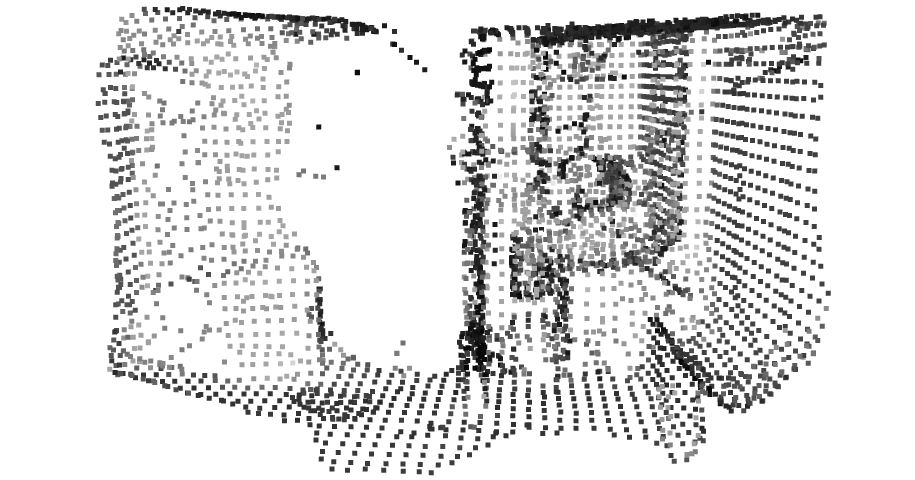}
    \end{minipage}
    \hfill
    \begin{minipage}[t]{0.48\linewidth}
        \centering
        \includegraphics[width=\linewidth]{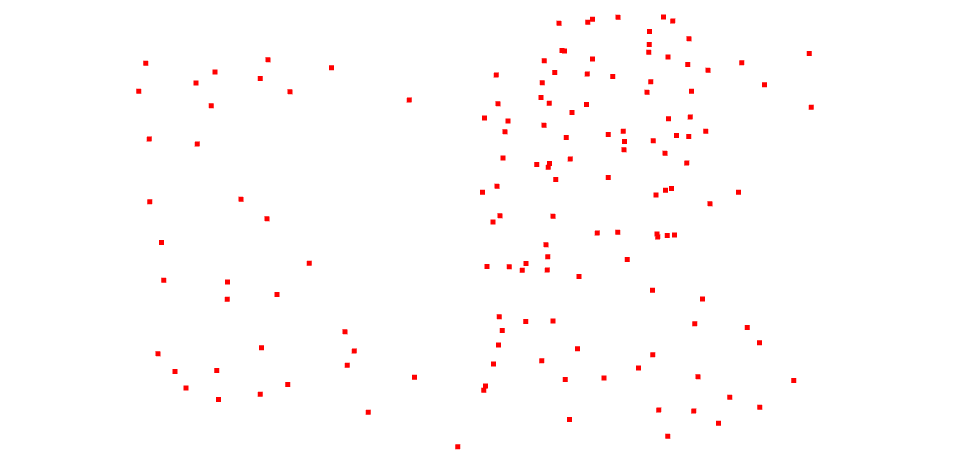}
    \end{minipage}
    \caption{Privacy effect: raw down-sampled cloud (left) reveals room layout, whereas ISS keypoints (right) hide many details while remaining matchable.}
    \label{fig:privacy}
\end{figure}

Figure~\ref{fig:privacy} illustrates how ISS-RegAuth provides better privacy protection compared to the baseline.
The raw down-sampled point cloud (left) reveals detailed room features—for instance, the distinctive curved shape clearly indicates the presence of a bathtub.
In contrast, ISS-RegAuth (right) retains only 1--2\% of structurally salient keypoints, making such objects much less recognizable at first glance while still preserving sufficient geometric information for reliable matching.
This sparsity significantly reduces the risk of privacy violations.

However, comprehensive privacy evaluation remains an open challenge.
We must first establish what constitutes privacy-sensitive information in point cloud data across various contexts (e.g., religious objects, medical equipment, lifestyle indicators), then assess the risk of privacy violation.
Developing quantitative privacy protection metrics and guidelines such as ISO/IEC 24745\cite{iso} is essential.

\subsection{Applications to Account Recovery}
A key advantage of indoor space authentication is its device-independent nature: the authentication credential is tied to a physical location rather than a specific device.
This makes it particularly well-suited for account recovery when users lose all trusted devices.

Conventional recovery mechanisms rely on secondary channels like email or SMS, which are vulnerable to phishing and SIM-swap attacks, or on hardware tokens that may be lost alongside other devices.
In contrast, indoor space authentication enables users to regain access by rescanning a pre-registered trusted space (e.g., home, office) using any LiDAR-capable device.
This approach provides an alternative recovery option that reduces dependence on third-party channels while preserving privacy.

\section{Conclusion and Future Work}
This paper presented ISS-RegAuth, which achieves superior authentication accuracy while reducing privacy exposure by extracting only 1–2\% of salient keypoints from LiDAR point clouds.
By resolving the planar surface dominance problem in prior methods, this device-independent approach shows promise for account recovery applications where users can regain access by rescanning their physical spaces.

Nevertheless, substantial research challenges remain.
In addition to comprehensive privacy evaluation, environmental robustness requires evaluation using longitudinal datasets that capture temporal changes (e.g., furniture rearrangement, seasonal decorations) and adversarial conditions.
While ISS-RegAuth demonstrates promising results, substantial further work is needed to realize indoor space authentication in real-world settings.

\end{document}